# Reflection of a few-cycle laser pulse on a metal nano-layer: generation of phase dependent wake-fields

Sándor Varró

*Research Institute for Solid State Physics and Optics*
*Letters : H-1525 Budapest, POBox 49, Hungary*
*E-mail : varro@sunserv.kfki.hu*

**Abstract.** The reflection and transmission of a few-cycle femtosecond Ti:Sa laser pulse impinging on a metal nano-layer have been analysed. The thickness of the layer was assumed to be of the order of 2-10 nm, and the metallic free electrons were represented by a surface current density at the plane boundary of a dielectric substrate. The target studied this way can be imagined for instance as a semi-transparent mirror produced by evaporating a thin aluminum layer on the surface of a glas plate. The exact analytical solution has been given for the system of the coupled Maxwell-Lorentz equations describing the dynamics of the surface current and the scattered radiation field. It has been shown that in general a non-oscillatory frozen-in wake-field appears following the main pulse with an exponential decay and with a definite sign of the electric field. The characteristic time of these wake-fields is inversely proportional with the square of the plasma frequency and with the thickness of the metal nano-layer, and can larger then the original pulse duration. The magnitude of these wake-fields is proportional with the incoming field strength, and the definite sign of them is governed by the cosine of the carrier-envelope phase difference of the incoming ultrashort laser pulse. As a consequence, when we let such a wake-field excite the electrons of a secondary target (say an electron beam, a metal plate or a gas jet), we obtain 100 percent modulation depth in the electron signal in a given direction. This scheeme can perhaps serve as a basis for the construction of a robust linear carrier-envelope phase difference meter.

## 1. Introduction

　　　The study of the interaction of intense few-cycle laser pulses with matter has brought a new, important branch of investigations in nonlinear optics, as Brabec and Krausz [1] emphasized in their review paper. The effect of the absolute phase (the carrier-envelope phase difference, in short: CE phase) on the nonlinear response of atoms and of solids interacting with a very short, few-cycle strong laser pulse has recently drawn considerable attention and has initiated a wide-spreading theoretical and experimental research. For instance Paulus et al. [2] have detected an anticorrelation in the shot-to-shot analysis of the photoelectron yield of ionization measured by two opposing detectors. This effect comes from the random variation of the CE phase (hence the direction and the magnitude of the electric field of the laser) from one pulse to the other. Such extreme short pulses can be used to monitor the details of photoelectron dynamics [3] or atomic inner-shell relaxation processes, like the Auger effect [4]. Concerning theory, the CE pase-dependence of the spatial asymmetry in photoionization has been extensively investigated by Chelkowski et al. [5], and by Milosevic et al. [6-7]. In the meantime the problem of the stabilization of the CE phase in the few-cycle laser pulse trains has been achieved [8-9]. On the basis of a simulation, using the time-dependent density functional approach, Lemell et al. [10] predicted the CE phase-dependence of the photoelectron yield in case of the surface photoelectric effect of metals in the tunnel regime. Apolonskiy et al. [11] and Dombi et. al. [12] have reported the measurement of this effect, but the absolut phase-dependence had a considerably smaller modulation (CE phase-sensitivity) in their experiment than predicted by the simulation. Fortier et al. has recently demonstrated the CE phase effect in quantum interference of injected photocurrents in semiconductors [13]. In the multiphoton regime Nakajima and Watanabe [14] has found theoretically CE phase effects in the bound state population of a Cs atom excited by nearly single-cycle pulses.

At this point we would like to note that there is a wide-spread opinion among researchers investigating the CE phase effects that these effects appear exclusively in *nonlinear* processes. In fact, as Fearn and Lamb [15] have shown already in 1991, the sine or the cosine character of the laser pulse make a difference in the *linear* photoionization dynamics if one takes into account the counter-rotating term in the interaction. As they wrote in Section IV. of their paper: "This suggests that…the time delay [of the electron signal] could be used to measure the phase of the field." A simple illustration of the linear CE phase effect has recently been considered by Ristow [16] in the case of a harmonic oscillator.

We have seen above, that in the theoretical works exclusively nonlinear quantum processes (photoionization, surface photoelectric effect) have been considered. In the present paper we briefly describe our theoretical analysis on the reflection and transmission of a few-cycle laser pulse on a thin metal layer represented by a surface current. Our analysis here, as in our earlier study [17], is based completely on classical electrodynamics and mechanics, in the frame of which we solve the system of coupled Maxwell-Lorentz equations of the incoming and scattered radiation and the classical surface current representing the metallic electrons.

In Section 2 we present the basic equations describing our model, and present the exact analytical solution of the scattering problem in the non-relativistic regime. Here we shall briefly analyse the exact solutions in the frequency domain. In Section 3 we discusse the temporal behaviour of the reflected (and transmitted) signal, and we shall show that a "pulse-decompression" and "freezing-in" of the radiation field may happen, yielding to the appearance of a quasi-static wake-field in the scattered signal.

## 2. The basic equations of the model and the exact analytic solutions in the frequency domain

The model to be used here we have already studied in our earlier work [17]. For completeness of the present paper, let us briefly summarize first the basic notations and equations, which can also be found in Ref. [17]. We take the coordinate system such that the first dielectric with index of refraction $n_1$ fills the region $z > l_2/2$, this is called region 1. In region 2 we place the thin metal layer of thickness $l_2$ perpendicular to the z-axis and defined by the relation $-l_2/2 < z < +l_2/2$. Region 3, $z < -l_2/2$, is assumed to be filled by the second dielectric having the index of refraction $n_3$. The thickness $l_2$ is assumed to be much smaller then the average skin depth of the incoming radiation. The target defined this way can be imagined as a thin metal layer evaporated, for instance, on a glass substrate. This layer, in fact is assumed to be represented by a sheet of electrons bound to region 2 and moving freely in the x-y plane. In case of perpendicular incidence the light would come from the positive z-direction, and it would be transmitted in the negative z-direction into region 3. The plane of incidence is defined as the y-z plane and the initial $\bar{k}$-vector is assumed to make an angle $\theta_1$ with the z-axis. In case of an s-polarized incoming TE wave the components of the electric field and the magnetic induction read $(E_x,0,0)$ and $(0, B_y, B_z)$, respectively. They satisfy the Maxwell equations

$$\partial_y B_z - \partial_z B_y = \partial_0 \varepsilon E_x, \quad \partial_z E_x = -\partial_0 B_y, \quad -\partial_y E_x = -\partial_0 B_z \quad , \tag{1}$$

where $\varepsilon = n^2$ is the dielectric constant and $n$ is the index of refraction. If we make the replacements $\varepsilon E_x \to -B_x$, $B_z \to E_z$ and $B_y \to E_y$ then we have the field components of a p-polarized TM wave $(0, E_y, E_z)$ and $(B_x, 0, 0)$, and we receive the following equations

$$\partial_z B_x = \partial_0 \varepsilon E_y, \quad -\partial_y B_x = \partial_0 \varepsilon E_z, \quad \partial_y E_z - \partial_z E_y = -\partial_0 B_x \quad . \tag{2}$$

In the followings we will consider only the latter case, namely the scattering of a p-polarized TM radiation field. From Eq. (2) we deduce that $B_x$ satisfies the wave equation, and in region 1 we take it as a superposition of the incoming plane wave pulse $F$ and an unknown reflected plane wave $f_1$

$$B_{x1} = F - f_1 = F[t - n_1(y\sin\theta_1 - z\cos\theta_1)/c] - f_1[t - n_1(y\sin\theta_1 + z\cos\theta_1)/c]. \tag{3}$$

From Eq.(2) we can express the components $E_y$ and $E_z$ of the electric field strength by taking into account Eq. (3)

$$E_{y1} = (\cos\theta_1/n_1)(F + f_1), \quad E_{z1} = (\sin\theta_1/n_1)(F - f_1) \quad . \tag{4}$$

In region 3 the general form of the magnetic induction $B_{x3}$ is the by now unknown refracted wave $g_3$

$$B_{x3} = g_3 = g_3[t - n_3(y\sin\theta_3 - z\cos\theta_3)/c]. \tag{5}$$

The corresponding components of the electric field strength are expressed from the above equation with the help of the first two equation of Eq. (2)

$$E_{y3} = (\cos\theta_3/n_3)g_3, \quad E_{z3} = (\sin\theta_3/n_3)g_3. \tag{6}$$

In region 2 the relevant Maxwell equations with the current density $\vec{j}$ read

$$\partial_z B_x = (4\pi/c)j_{y2} + \partial_0 \varepsilon E_y, \quad \partial_y E_z - \partial_z E_y = -\partial_0 B_x. \tag{7}$$

By integrating the two equations in Eq. (7) with respect to z from $-l_2/2$ to $+l_2/2$ and taking the limit $l_2 \to 0$, we obtain the boundary conditions for the field components

$$[B_{x1} - B_{x3}]_{z=0} = (4\pi/c)K_{y2}, \quad [E_{y1} - E_{y3}]_{z=0} = 0, \tag{8}$$

where $K_{y2}$ is the y-component of the surface current in region 2. This surface current can be expressed in terms of the local velocity of the electrons in the metal layer

$$K_{y2} = e(d\delta_y/dt)l_2 n_e, \quad (4\pi/2c)K_{y2} = (m/e)\Gamma(d\delta_y/dt), \tag{9}$$

$$\Gamma \equiv 2\pi(e^2/mc)l_2 n_e, \quad \Gamma = (\omega_p/\omega_0)^2(\pi l_2/\lambda_0)\omega_0, \quad \kappa \equiv \Gamma/\omega_0 = \pi(\omega_p/\omega_0)^2(l_2/\lambda_0) \quad , \tag{10}$$

where for later convenience we have introduced $\omega_0$, $\lambda_0 = 2\pi c/\omega_0$, the carrier frequency and the central wavelength of the incoming light pulse, and $n_e, \omega_p = \sqrt{4\pi n_e e^2/m}$ denote the density of electrons and the corresponding plasma frequency in the metal layer, respectively. In Eq. (9) $\delta_y$ denotes the local displacement of the electrons in the metal layer for which we later write down the Lorentz equation (Newton equation in the non-relativistic regime) in the presence of the complete electric field. We remark that in reality the thickness $l_2$ is, of course, not infinitesimally small, rather, it has a finite value which is anyway assumed to be smaller then the skin depth $\delta_{skin} = c/\sqrt{\omega_p^2 - \omega_0^2}$. In order to have a feeling on the size of the parameters coming into our analysis, let us take some illustrative examples. For instance, for $n_e = 10^{22} / cm^3$, $\delta_{skin} = \lambda_0 / 14$, and for $n_e = 10^{23} / cm^3$, $\delta_{skin} = \lambda_0 / 47$, where we have taken $\lambda_0 = 800 nm$ and $\omega_0 = 2.36 \times 10^{15} s^{-1}$ for a Ti:Sa laser. For the damping parameter $\Gamma$ in the first case, if we take $l_2 = \lambda_0 / 400 = 2nm$, we have $\kappa \equiv \Gamma / \omega_0 = 1/20$. In the second case, for the same thickness $l_2 = \lambda_0 / 400 = 2nm$ we have $\kappa \equiv \Gamma / \omega_0 = 0.18$.

From Eq. (8) with the help of Eq. (9) we can express $f_1$ and $g_3$ in terms of $\delta'_y(t')$

$$f_1(t') = (1/(c_1 + c_3))[(c_3 - c_1)F(t') - 2c_3(m/e)\Gamma\delta'_y(t')], \tag{11}$$

$$g_3(t') = (2c_1/(c_1 + c_3))[F(t') - (m/e)\Gamma\delta'_y(t')], \tag{12}$$

where the prime on $\delta_y$ denotes the derivative with respect to the retarded time $t' = t - yn_1 \sin\theta_1 / c$ which is equal to $t - yn_3 \sin\theta_3 / c$, securing Snell's law of refraction $n_1 \sin\theta_1 = n_3 \sin\theta_3$ to hold. Moreover, in Eqs. (11) and (12) we have introduced the notations $c_1 = \cos\theta_1 / n_1$, $c_3 = \cos\theta_3 / n_3$. We would like to emphasize that Eqs. (11) and (12) are valid in complete generality, that is, they hold for both non-relativistic and relativistic kinematics of the local electron displacement $\delta_y(t')$. For an interaction with a TM wave this displacement is uniform (along lines of constant *x*-values) in the direction perpendicular to the plane of incidence (the *y-z* plane), so it does not depend on the *x*-coordinate. As the incoming wave impinges on the surface at region 2 its (plane) wave fronts sweep this surface creating a superluminar polarization wave, described by the local displacement $\delta_y(t')$ of the electrons. Because of the continuity of $E_y$, Eq. (8), in the Newton equation for the displacement of the electrons in the surface current we can use for instance the force term $E_{y1} = c_1(F + f_1)$ according to Eq. (4), and neglect the magnetic induction. By taking Eq. (11) also into account we have

$$\delta''_y(t') = b[(e/m)F(t') - \Gamma\delta'_y(t')], \tag{13}$$

$$b \equiv 2c_1c_3/(c_1 + c_3), \quad c_1 \equiv \cos\theta_1 / n_1, \quad c_3 \equiv \cos\theta_3 / n_3 = (1/n_3^2)\sqrt{n_3^2 - n_1^2(1 - \cos^2\theta_1)}. \tag{13a}$$

In obtaining the last formula in Eq. (13a) we have used Snell's law of refraction mentioned already a couple of lines before. For definiteness, we impose the initial conditions on the electron displacement $\delta_y(-\infty) = 0$ and $\delta'_y(-\infty) = 0$. Owing to Eqs. (11) and (12) the solution of Eq. (13) gives at the same time the complete solution of the scattering problem. We see that the equation of motion for the local displacement $\delta_y(t')$ contains a damping term with a damping parameter $b\Gamma$ where $\Gamma$ has been defined in Eq. (10). This latter constant is proportional with the squared plasma frequency and the thickness of the electron layer. The appearance of the damping term is a manifestation of the radiation reaction coming formally from the boudary conditions in the present description. Since $\Gamma$ is proportional with the electron density, this effect is due to the collective response of the electrons to the action of the complete (not only the incoming) radiation field, which, on the other had reacts back to the electrons. In the present description it is not possible to divide into steps these "consecutive" effects, as in the usual treatments of the radiation back-reaction. Eq. (13) can be solved exactly for an arbitrary incoming field $F(t)$. By calculating the Fourier transforms of Eqs. (11), (12) and (13) we can give an exact solution of the scattering problem in the frquency space, namely

$$\tilde{f}_1(\omega) = -\frac{\tilde{F}(\omega)}{b\Gamma - i\omega}\left[b\Gamma + \frac{c_3 - c_1}{c_3 + c_1}i\omega\right] = -\frac{\tilde{F}(\omega)}{b\kappa - iv}\left[b\kappa + \frac{c_3 - c_1}{c_3 + c_1}iv\right], \tag{15}$$

$$\tilde{g}_3(\omega) = -\frac{2c_1}{c_1 + c_3}\frac{i\omega\tilde{F}(\omega)}{b\Gamma - i\omega} = -\frac{2c_1}{c_1 + c_3}\frac{iv\tilde{F}(\omega)}{b\kappa - iv}, \tag{16}$$

where we are using the dimensionless quantities $\kappa$ and $\nu$ given by the definitions
$$\kappa \equiv \Gamma/\omega_0, \quad \nu \equiv \omega/\omega_0, \tag{17}$$
and the geometrical factor $b$ was defined in Eq. (13a). It can be proved that the Fourier components of the reflected and the transmitted fluxes (to be calculated from Eqs. (15) and (16)) satisfy the following sum rule
$$c_1|\tilde{f}_1(\omega)|^2 + c_3|\tilde{g}_3(\omega)|^2 = c_1|\tilde{F}(\omega)|^2. \tag{18}$$

By now we have not specified the explicit form of the incoming field. Let us assume that it is a Gaussian quasi-monochromatic field with a carrier frequency $\omega_0$ having the carrier-envelope phase difference (CE phase) $\varphi$,
$$F(t) = F_0 \exp(-t^2/2\tau^2)\cos(\omega_0 t + \varphi), \tag{19}$$
where $\tau = \tau_L/2$ with $\tau_L$ being the full temporal width of the pulse's intensity. The Fourier transform of the incoming pulse given by Eq. (19) reads
$$\tilde{F}(\omega) \equiv \int_{-\infty}^{\infty} dt F(t) e^{i\omega t} = F_0 \tau (\pi/2)^{1/2} \exp[-\tau^2(\omega^2 + \omega_0^2)/2] \cdot \left( e^{i\varphi} e^{-\omega\omega_0\tau^2} + e^{-i\varphi} e^{+\omega\omega_0\tau^2} \right), \tag{20a}$$
and its modulus squared is the following $\pi$-periodic function of the CE phase
$$|\tilde{F}(\omega)|^2 = 2\{F_0 \tau(\pi/2)^{1/2}\exp[-\tau^2(\omega^2+\omega_0^2)/2]\}^2 (\cosh 2\tau^2\omega\omega_0 + \cos 2\varphi). \tag{20b}$$
From Eq. (20b) it is immediately seen that for any linear excitation process (whose response function is proportional with the incoming intensity) the frequency dependence of the modulation function ("visibility" or "contrast function") of the response function is given by the expression
$$M(\omega) = \frac{I_{\max}(\omega) - I_{\min}(\omega)}{I_{\max}(\omega) + I_{\min}(\omega)} = \frac{|\tilde{F}_{\max}(\omega)|^2 - |\tilde{F}_{\min}(\omega)|^2}{|\tilde{F}_{\max}(\omega)|^2 + |\tilde{F}_{\min}(\omega)|^2} = \frac{1}{\cosh 2\tau^2\omega\omega_0}, \tag{21}$$
where $I_{\max, \min}(\omega)$ here may mean both the maximum (minimum) values of the reflected flux $c_1|\tilde{f}_1(\omega)|^2$ and the refracted flux $c_3|\tilde{g}_3(\omega)|^2$ at a particular frequency $\omega$, as we vary the CE phase. This means that as we vary the CE phase between 0 and $\pi$, the modulation depth is given by Eq. (21). It is clear that if $\omega_0\tau$ is very large (which is the case of many-cycle pulses), then the modulation function is practically zero. On the other hand, when $\omega_0\tau$ is not large (which is the case of few-cycle pulses), then for small frequencies ($\omega\tau \ll 1$) we can have a modulation close to 100%. From Eqs. (15), (16) and (21) one obtains that the modulation of both the reflected and the transmitted signal is given by $1/\cosh 2\tau^2\omega\omega_0$ at a particular frequency. On the other hand neither of the reflection coefficient nor the transmission coefficient depend on the CE phase in the linear regime. It is clear that if the response function contains a resonance at a small frequency ($\omega\tau \ll 1$), then there is a better chance to observe the mentioned linear CE-phase-dependence at that particular resonance frequency.

At this point let us note that Eqs. (15) and (16) are more general then the original relations, Eqs. (11), (12) and (13), written in the time domain, where the indeces of refractions $n_1$ and $n_3$ have been taken as mere constants. The optional dispersion can be taken into account in Eqs. (15) and (16) by putting *by hand* a frequency dependence into $n_1$ and $n_3$. In the present paper we are not dealing with this aspect of the problem. Our description is approximate to the real scattering process in an other respect, too, namely it relies on the plane-wave model of the incoming, reflected and the refracted field. This is a standard approximation which is used in the texbooks throughout. Equations (15) and (16) can be superimposed for an assembly of incoming plane waves of different propagation direction, and this description would be suitable to handle the scattering of a *beam* with a limited transverse extension. In the present paper we are not dealing with this aspect of the scattering problem, either. Thus, our results are suitable to describe the scattering of unfocused beams.

## 3. The appearance of frozen-in wake-fields in the scattered signal

As one sees from Eqs. (15) and (16) both the reflected and the refracted Fourier components have a pole at $\omega = -ib\Gamma$, which means that in the time domain a functional dependence of the form $\propto \exp(-b\Gamma t)$ is expected with a decay time $T/2\pi b\kappa$, where $b$ and $\kappa$ were defined in Eqs. (13a) and (17), respectively, and $T = 2\pi/\omega_0$ is the central period of the incoming field. This can easily be shown by integrating the ordinary first order differential equation Eq. (13) for the local velocity $\delta'_y(t')$, which determines through Eq. (11) and (12) the scattered fields,

$$\delta'_y(t') = e^{-b\Gamma t'} \int_{-\infty}^{t'} du [b(e/m)F(u)] e^{+b\Gamma u} . \tag{22}$$

The integral in Eq. (22) can be easily evaluated for any usual functional form of the incoming field. From Eq. (11) and (22) the completete reflected field can be obtained,

$$f_1(t') = \frac{c_3 - c_1}{c_3 + c_1} F(t') - 2\pi b^2 \kappa \exp[-2\pi b\kappa(t'/T)] \int_{-\infty}^{t'} d(u/T) F(u) \exp[+2\pi b\kappa(u/T)], \tag{23a}$$

where now in region 1 the retarded time reads
$$t' = t - n_1(y\sin\theta_1 + z\cos\theta_1)/c . \tag{23b}$$

For a model pulse of a $\sin^2$ envelope of finite support (which has been widely used in numerical simulations), the integral can be calculated analytically. Henceforth, in our illustrative numerical examples we will use the Gaussian pulse defined in Eq. (19).

Before entering into some numerical examples of the temporal behaviour of the scattered field, let us make some general remarks. It is clear that if the pulse duration of the incoming field $F(t)$ is much smaller than the characteristic time $1/b\Gamma = T/2\pi b\kappa$ of the exponential factor, then $F(t)$ cuts off a small region in the inegration interval in Eq. (22) because of the relative smoothness of the exponential function. This means that the upper limit of the integral can be extended up to infinity within a reasonable approximation. Hence the *asympthotic behaviour of the velocity* $\delta'_y(t')$ can certainly be well represented by the approximate formula

$$\delta'_y(t') \approx e^{-b\Gamma t'} \int_{-\infty}^{+\infty} du [b(e/m)F(u)] e^{+b\Gamma u} = e^{-b\Gamma t'} b(e/m) \tilde{F}(-ib\Gamma) , \tag{24a}$$

where $\tilde{F}(-ib\Gamma)$ denotes the Fourier transform of the incoming field. By taking Eq. (20a) into account, we obtain from Eq. (23a)

$$\delta'_y(t') \approx e^{-b\Gamma t'} b(e/m)\left(F_0 \tau\sqrt{2\pi}\right) \exp\left[-(\omega_0^2 \tau^2/2)(1 - b^2\kappa^2)\right] \cos(\varphi + \omega_0^2 \tau^2 b\kappa), \tag{24b}$$

$$\delta'_y(t')/c$$
$$\approx -\left(b\omega_0 \tau\sqrt{2\pi}\right) \exp\left[-(\omega_0^2 \tau^2/2)(1 - b^2\kappa^2)\right] \cdot \mu \exp\left(-2\pi b\kappa \frac{t'}{T}\right) \cos(\varphi + \omega_0^2 \tau^2 b\kappa), \tag{24c}$$

where we have introduced the usual intensity parameter $\mu$ with the definition

$$\mu \equiv \frac{eF_0}{mc\omega_0} = 10^{-9}\sqrt{I}/E_{ph}, \quad \mu^2 = 10^{-18} I\lambda^2 . \tag{24d}$$

In the numerical expressions in Eq. (24d) $I$ denotes the mean intensity of the pulse measured in W/cm$^2$, $E_{ph}$ is the mean photon energy in eV-s, and $\lambda$ denotes the central wavelength of the pulse measured in microns (we have not displayed the numerical prefactors of order unity). According to Eq. (24c), the polarization current and (due to Eq. (11)) the scattered field itself contains a *wake-field* which is a *frozen-in non-oscillatory quasi-static field* propagating in the direction $(0, \sin\theta_1, \cos\theta_1)$ after leaving the metal layer. The decay time of this wake-field is given by $T/2\pi b\kappa$, which can be much larger than an optical period, as is shown on **Fig. 1**, where we have taken $\kappa = 1/20$. Since we are dealing just with few-cycle incoming pulses, this means, that the wake-field is still present, after the main pulse has passed. Moreover, from Eq. (23a) it can be seen, that at the Brewster angle of incidence ($c_1 = c_3 \to \theta \approx 56°$ if $n_1 = 1$, $n_3 = 1.5$) only the wake-field is propagating in that particular direction. It is remarkable, that, as can be seen from Eqs. (24b-c), the amplitude of the wake-field is proportional with the cosine of the CE phase $\varphi$. Thus, for instance, by varying the CE phase, the sign of the quasi-static wake-field can be reversed. On **Fig. 2** an illustrative example is shown for the temporal behavior of the reflected signal at Brewster angle of incidence of an incoming one-cycle Ti:Sa laser pulse.

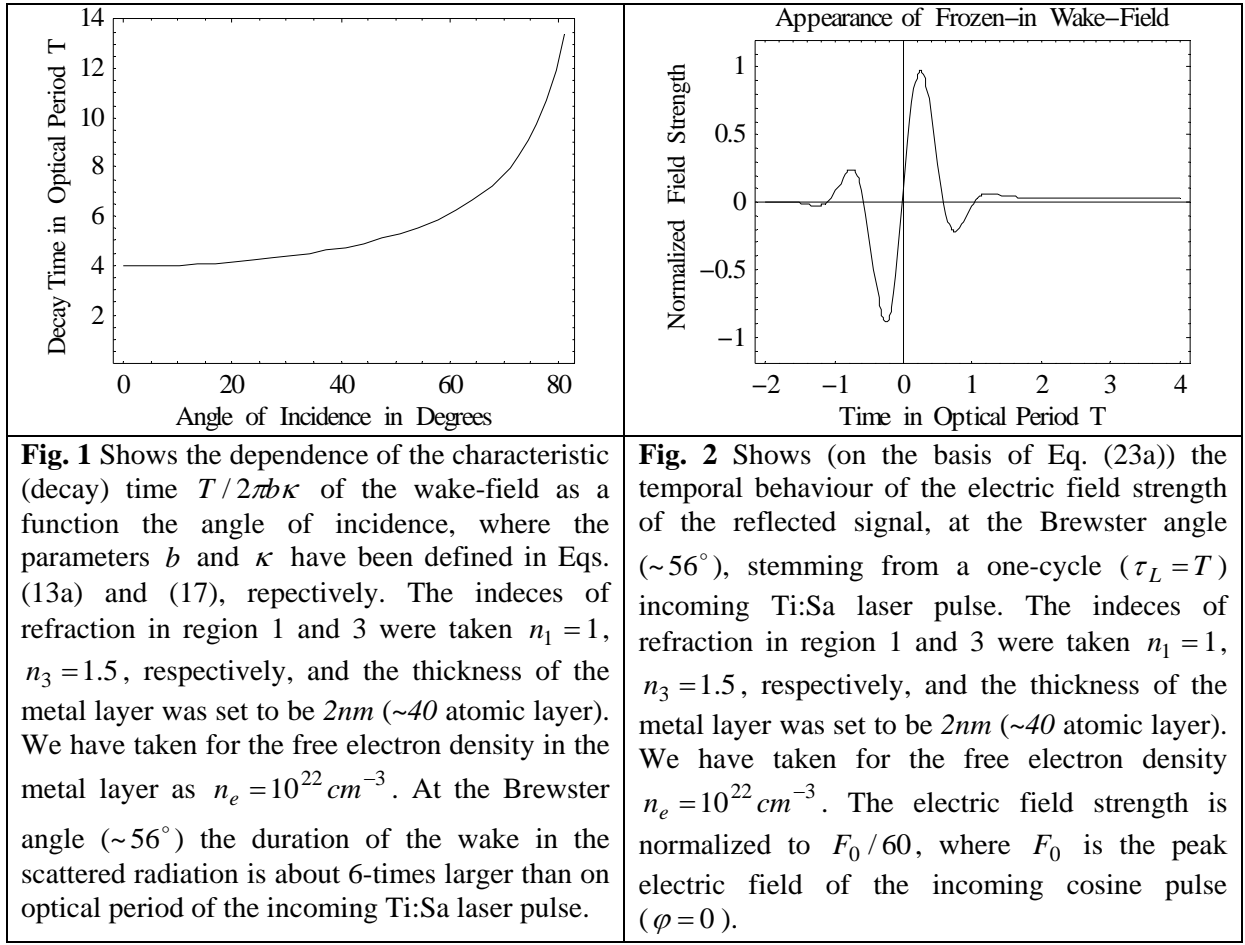

**Fig. 1** Shows the dependence of the characteristic (decay) time $T/2\pi b\kappa$ of the wake-field as a function the angle of incidence, where the parameters $b$ and $\kappa$ have been defined in Eqs. (13a) and (17), repectively. The indeces of refraction in region 1 and 3 were taken $n_1 = 1$, $n_3 = 1.5$, respectively, and the thickness of the metal layer was set to be *2nm* (~*40* atomic layer). We have taken for the free electron density in the metal layer as $n_e = 10^{22} cm^{-3}$. At the Brewster angle (~56°) the duration of the wake in the scattered radiation is about 6-times larger than on optical period of the incoming Ti:Sa laser pulse.

**Fig. 2** Shows (on the basis of Eq. (23a)) the temporal behaviour of the electric field strength of the reflected signal, at the Brewster angle (~56°), stemming from a one-cycle ($\tau_L = T$) incoming Ti:Sa laser pulse. The indeces of refraction in region 1 and 3 were taken $n_1 = 1$, $n_3 = 1.5$, respectively, and the thickness of the metal layer was set to be *2nm* (~*40* atomic layer). We have taken for the free electron density $n_e = 10^{22} cm^{-3}$. The electric field strength is normalized to $F_0/60$, where $F_0$ is the peak electric field of the incoming cosine pulse ($\varphi = 0$).

From **Fig. 2** we see that the magnitude of the electric field of the wake-field is about 600 times smaller than the peak field strength of the incoming cosine pulse. The peak electric field strength $F_0$ of a laser field can be calculated from the laser intensity $I_0$ according to the following formula,

$$[F_0/(V/cm)] = 27.46 \times [I_0/(W/cm^2)]^{1/2} . \tag{25}$$

Thus, for an incoming laser of intensity $I_0 \approx 10^{12} W/cm^2$, the amplitude of the frozen-in wake-field in the reflected signal is of oder of 50000 *V/cm*, which is quite a large value for a quasi-static field. As we have already mentioned above the amplitude and sign of this wake-field can be varied by changing the CE phase of the incoming pulse. The idea naturally emerges, that when we let such a wake-field excite the electrons of a secondary target  – say an electron beam, a second metal plate or a gas jet – *we may obtain 100 percent modulation depth in the electron signal (acceleration, tunnel-ionization) in a given direction, even for a relatively low intensity incoming field.* This scheeme can perhaps serve as a basis for the construction of a robust linear carrier-envelope phase difference meter. At the end of the present section we would like to emphasize that the generation of the wake-fields discussed above is a linear process, since the amplitude of the wake-field is simply proportional with the amplitude of the incoming field. This means that for the observation of the effect considered here one does not need extremely high laser intensities.

**Acknowledgements.** This work has been supported by the Hungarian National Scientific Research Foundation, OTKA, Grant No. T048324.